\documentclass[aps,showpacs,floatfix,twocolumn,byrevtex,superscriptaddress]{revtex4-1}

\usepackage{amsmath}
\usepackage{amssymb}
\usepackage{amstext}
\usepackage{amsopn}
\usepackage{amsfonts}
\usepackage{amsxtra}
\usepackage[english]{babel}
\usepackage{amsmath}
\usepackage{graphicx}
\usepackage{float}
\usepackage{bm}
\usepackage{multirow}
\usepackage{dcolumn}
\usepackage{hyperref}
\usepackage{CJK}

\begin{document}

\title{Observation of an Emergent Coherent State in Iron-Based Superconductor KFe$_\mathbf{2}$As$_\mathbf{2}$}
\author{Run Yang}
\affiliation{Beijing National Laboratory for Condensed Matter Physics, Institute of Physics,
  Chinese Academy of Sciences, Beijing 100190, China}
\affiliation{Condensed Matter Physics and Materials Science Division, Brookhaven National Laboratory,
  Upon, New York 11973, USA}
\affiliation{School of Physical Sciences, University of Chinese Academy of Sciences, Beijing 100049, China}
\author{Zhiping Yin}
\email[]{yinzhiping@bnu.edu.cn}
\affiliation{Center for Advanced Quantum Studies and Department of Physics, Beijing Normal
  University, Beijing 100875, China}
\author{Yilin Wang}
\author{Yaomin Dai}
\author{Hu Miao}
\affiliation{Condensed Matter Physics and Materials Science Division, Brookhaven National Laboratory,
  Upon, New York 11973, USA}
\author{Bing Xu}
\affiliation{Beijing National Laboratory for Condensed Matter Physics, Institute of Physics,
  Chinese Academy of Sciences, Beijing 100190, China}
\author{Xianggang Qiu}
\email[]{xgqiu@iphy.ac.cn}
\affiliation{Beijing National Laboratory for Condensed Matter Physics, Institute of Physics,
  Chinese Academy of Sciences, Beijing 100190, China}
\affiliation{School of Physical Sciences, University of Chinese Academy of Sciences, Beijing 100049, China}
\affiliation{Collaborative Innovation Center of Quantum Matter, Beijing 100084, China}
\author{Christopher C. Homes}
\email[]{homes@bnl.gov}
\affiliation{Condensed Matter Physics and Materials Science Division, Brookhaven National Laboratory,
  Upon, New York 11973, USA}
\date{\today}
%
%

\begin{abstract}
The optical properties of KFe$_2$As$_2$ have been measured for light polarized in the
\emph{a-b} planes over a wide temperature and frequency range.  Below $T^\ast\simeq 155$~K,
where this material undergoes an incoherent-coherent crossover, we observed a new coherent
response emerging in the optical conductivity.  A spectral weight analysis suggests that
this new feature arises out of high-energy bound states.  Below about $T_{\rm FL} \simeq 75$~K
the scattering rate for this new feature is quadratic in temperature, indicating a
Fermi-liquid response.  Theory calculations suggest this crossover is dominated by
the $d_{xy}$ orbital. Our results advocate for Kondo-type screening as the mechanism for
the orbital-selective incoherent-coherent crossover in hole-overdoped KFe$_2$As$_2$.
%
%
\end{abstract}


\maketitle

%
%
%
Investigating the role of electronic correlations in systems with large orbital degrees
of freedom remains a key challenge in understanding the multi-orbital physics in iron-based
superconductors (FeSCs)~\cite{Richard2009,Yin2011,Qazilbash2009,Georges2013,Yu2011,Lu2008,Chen2014}.
Previous investigations found that, due to Hund's rule coupling, the electron correlations
strongly depend on the band filling and are responsible for the electron-hole asymmetry in
FeSCs~\cite{DeMedici2014,DeMedici2011}.  Electron doping weakens the correlations and
finally results in Fermi-liquid (FL) behavior~\cite{Ning2010}, while the hole doping makes
the system more correlated~\cite{Hardy2016}, due to strong Hund's rule coupling and
closer proximity to half filling~\cite{DeMedici2011}.  KFe$_2$As$_2$, as an extremely
hole-doped FeSC, shows strong electronic correlations and bad-metal behavior at high
temperature.  However, upon cooling, the electrons start to form coherent quasiparticles
around $T^\ast\sim 155$~K, and shows FL behavior below $T_{\rm FL} \simeq 75$~K
\cite{Eilers2016,Hardy2013,Haule2009}; this has prompted an extensive debate between
the orbital-selective Mottness~\cite{DeMedici2014,Liu2015,Lu2008} and Hund's rule
coupling induced Kondo-type screening as the mechanism for the incoherent-coherent
crossover \cite{Fanfarillo2015,Backes2015,Miao2016,Stadler2015}.

In this Letter we examine the temperature dependence of the optical conductivity of
KFe$_2$As$_2$ to study the incoherent-coherent crossover. As the temperature is
lowered across $T^\ast$, we observe that spectral weight is transferred
from high ($\simeq 2000 - 3000$~cm$^{-1}$) to low ($\lesssim 1000$~cm$^{-1}$) energy region
into a new emergent Drude peak, which displays a FL behavior below $\simeq 75$~K.
To further investigate this behavior we performed dynamical mean field
theory (DMFT) calculations that suggest that the incoherent-coherent crossover is
governed by the $d_{xy}$ orbital.  The FL behavior indicates the Kondo-type
screening of local spin moments, which is consistent with susceptibility~\cite{Hardy2013} and
transport measurements~\cite{Xiang2016} and in agreement with the theoretical
prediction~\cite{Stadler2015,Aron2015}.  However, the absence of a Mott gap in the 
high-temperature optical spectra rules out the possibility of orbital-selective 
Mottness.   Thus, we propose a Kondo-type screening as the mechanism responsible 
for the orbital-selective incoherent-coherent crossover in KFe$_2$As$_2$, which 
is close to half filling.  This result introduces constraints for further 
theoretical investigations to understand the orbital physics as well as
the pairing mechanism in FeSCs.

%
%
%

High-quality single crystals of KFe$_2$As$_2$ with good cleavage planes (001) were
synthesized using self-flux method~\cite{Kim2011}.  The reflectance from freshly-cleaved
surfaces has been measured over a wide temperature ($\sim 4$ to 300~K) and frequency
range ($\sim 2$~meV to about 5~eV) at a near-normal angle of incidence for light polarized
in the \emph{ab}-planes using an \emph{in situ} evaporation technique~\cite{Homes93}
The optical conductivity has been determined from a Kramers-Kronig analysis of the
reflectivity (the reflectivity and the details of the Kramers-Kronig analysis may be
found in the Supplemental Information).

%
%

\begin{figure}[tb]
\centerline{
\includegraphics[width=0.90\columnwidth]{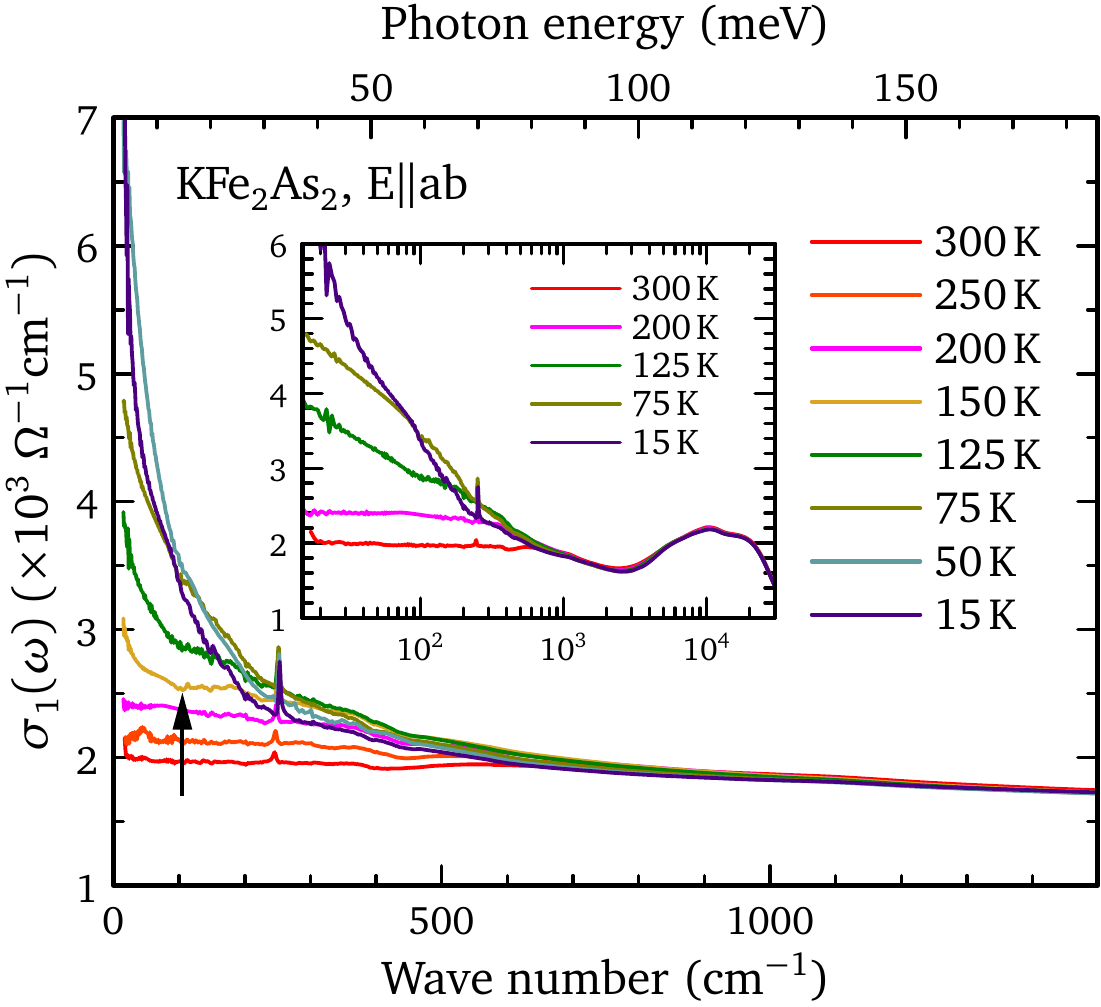}
}
\caption{The temperature dependence of the real part of the optical conductivity of
KFe$_2$As$_2$ above and below the incoherent-coherent crossover $T^\ast$.
Inset: The optical conductivity at several temperatures over a broad frequency range.}
\label{fig:sigma}
\end{figure}

The temperature dependence of the optical conductivity $\sigma_{1}(\omega)$ is
shown in Fig.~\ref{fig:sigma} in the infrared region.  The free carrier response
is typically a Lorentzian centered at zero frequency where the width at half
maximum is the scattering rate.  At high temperature, the conductivity is low
and resembles a bad metal; the nearly flat frequency response indicates a large
scattering rate, signaling an almost incoherent response.  As the temperature is
reduced, the low-frequency conductivity increases gradually until 150~K, at which
point a new Drude-like peak appears superimposed on the broad response, resulting
in a kink in the low-energy conductivity (denoted by the arrow in Fig.~\ref{fig:sigma},
as well as the sudden change in slope in the inset).  This new peak increases dramatically
in strength and narrows quickly with decreasing temperature, reflecting its small scattering
rate and coherent character~\cite{Dai2014}.

%
%
%
%

Because of the multi-band nature of FeSCs there is typically more than one type
of free carrier present.  Consequently, to analyze the optical conductivity we
employ a Drude-Lorentz model with multiple Drude components,
\begin{equation}
\sigma_{1}(\omega)=\frac{2\pi}{Z_{0}}\!\left[\sum_{j}\frac{\omega^{2}_{p,j}\tau_j}{(1+\omega^{2}\tau_{j}^{2})}+
  \sum_{k}\frac{\gamma_{k}\omega^{2}\Omega^{2}_{k}}{(\omega^2_{k}-\omega^2)^2+\gamma_{k}^{2}\omega^2}\right],
  \label{eq:dl}
\end{equation}
where $Z_{0}\simeq 377\,\Omega$ is the impedance of free space. The first term describes
a sum of free-carrier Drude responses with plasma frequencies $\omega_{p,j}^{2}=4\pi n_j e^{2}/m_j^{*}$
($n_j$ represents the carrier concentration and $m_j^{*}$ the effective mass), and scattering
rates $1/\tau_j$.
The second term is a sum of Lorentz oscillators with position $\omega_{k}$, width $\gamma_{k}$,
and oscillator strength $\Omega_{k}$.
At high temperature ($T> 150$~K), the optical conductivity is described quite well
by two Drude components and two Lorentz terms [Fig.~S2(a) of the Supplementary
Material]: a narrow Drude (D1) with a $T$-dependent scattering rate, a broad Drude (D2),
which is almost $T$ independent, indicating two groups of carriers~\cite{Wu2010,Dai2013}.
The narrow Drude reflects the coherent response and the broad Drude corresponds to
the incoherent background.  Below 150~K, the newly formed peak and corresponding kink
in $\sigma_1(\omega)$ makes it difficult to fit the low-energy response with only two Drude
components, so a third Drude component (D3) has been introduced.  The existence of a new
Drude component can also be realized by fitting the reflectivity and the imaginary part
of the optical conductivity [Table~I and Figs.~S2 and S3 in the Supplementary Material].

%
%
%
%
\begin{figure}[tb]
\centerline{
\includegraphics[width=0.9\columnwidth]{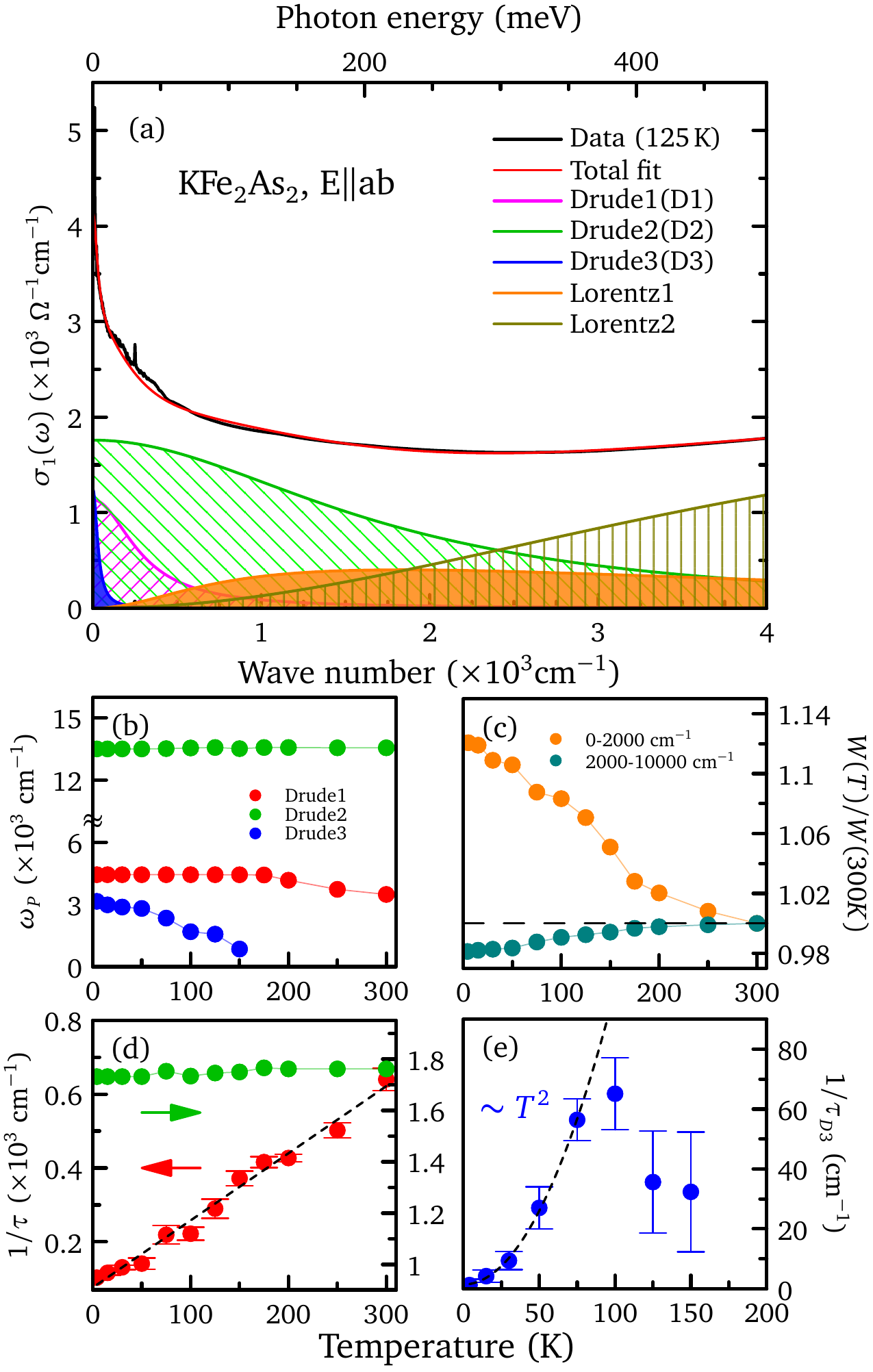}
}
\caption{(a) Fit of the Drude-Lorentz model to the $\sigma_1(\omega)$ of
KFe$_2$As$_2$ at 125~K, decomposed into individual Drude and Lorentz terms.
(b) The plasma frequency $\omega_{p}$ for the three Drude components.
(c) Temperature dependence of the spectral weight, $W(\omega_1,\omega_2,T)$,
for various lower and upper cutoff frequencies.
(d)Scattering rates for the narrow Drude (D1) (red), broad Drude (D2) (green) components
are extracted from the fits.  The dashed line is the linear fit to $1/\tau_{D1}$.
(e) The scattering rate of the emergent Drude component. The dashed line is
the quadratic fit to $1/\tau_{D3}$ below 75~K.}
\label{fig:fits}
\end{figure}

%
%

%
%
\begin{figure*}[tb]
\centerline{
\includegraphics[width=1.7\columnwidth]{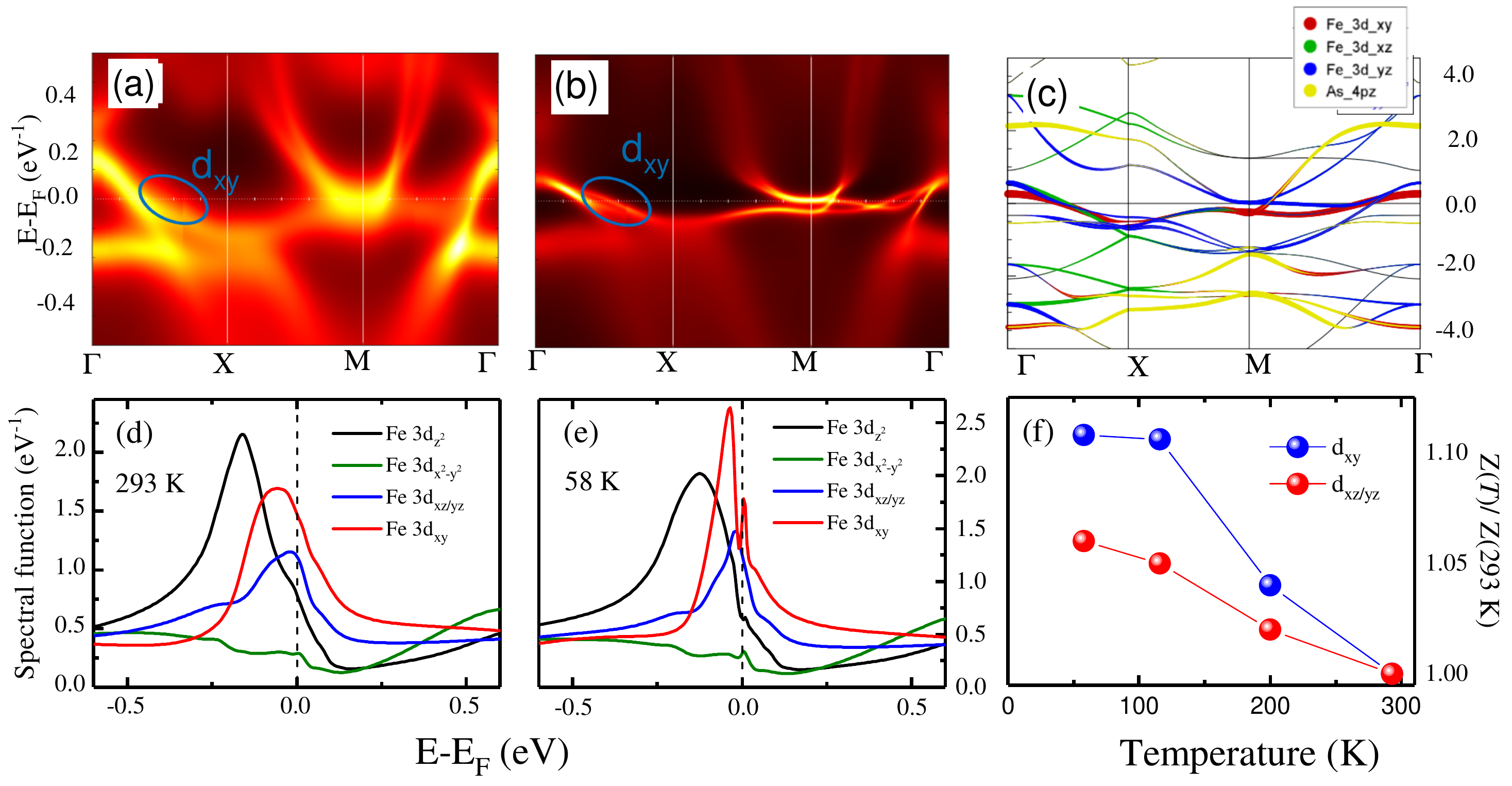}
}
\caption{Electronic structure of KFe$_2$As$_2$ above and below the incoherent-coherent
crossover.
(a) DFT+DMFT band structure for KFe$_2$As$_2$ at 293~K and (b) 58~K; the blue ellipses
denote the band of primarily $d_{xy}$ character.
(c) DFT band structure.
(d) The Fe-\emph{d} orbital character for the density of states of KFe$_2$As$_2$
at 293~K and (e) 58~K.
(f) The temperature-dependent quasiparticle spectral weight $Z$ calculate by
integrating the spectral function near $E_{\rm F}$.}
\label{fig:dmft}
\end{figure*}

Below $T^\ast$ the data is reproduced quite well with three Drude and two Lorentz
components, as illustrated by the fit to the data at 125~K in Fig.~\ref{fig:fits}(a).
The temperature dependence of the Drude components are summarized in Figs.~\ref{fig:fits}(b-e).
As the temperature is reduced, the values for $\omega_{p}$ for the narrow and broad Drude
terms, D1 and D2, respectively, remain essentially constant.  Below $T^\ast$ the plasma
frequency for newly developed Drude component (D3) increases steadily and $\omega_{p,D3}^{2}(T)$
follows a mean-field temperature dependence [Fig.~S4(a) in the Supplementary Material].
Figures~\ref{fig:fits}(d) and \ref{fig:fits}(e) show the temperature dependence
of the scattering rates; $1/\tau_{D2}$ is almost temperature independent, while
$1/\tau_{D1}\propto T$.  Below $T^\ast$, $1/\tau_{D3}$ initially increases
with temperature, but for $T < T_{\rm FL}$ FL behavior is observed, $1/\tau_{D3}
\propto T^{2}$.

%
%
In order to understand the origin of the new peak (D3), we calculate the
spectral weight of $\sigma_1(\omega)$,
\begin{equation}
  W(\omega_1,\omega_2,T)=\int_{\omega_1}^{\omega_2} \sigma_1(\omega,T)\,d\omega,
  \label{eq:weight}
\end{equation}
over different frequency intervals and normalize the result to the room-temperature value.
In Fig.~\ref{fig:fits}(c), we notice that, upon cooling, especially below 150~K, the
spectral weight between $\sim 0-2000$~cm$^{-1}$ is greatly enhanced, while the
spectral weight in the high-energy range ($2000-10\,000$~cm$^{-1}$) is suppressed.
Since $\omega_{p,D1}$ and $\omega_{p,D2}$ are temperature-independent below 150~K,
and the overall spectral weight up to $10\,000$~cm$^{-1}$ remains constant [Fig.~S3(a)]
(the optical conductivity above $10\,000$~cm$^{-1}$ does not vary with temperature),
we propose that the new Drude component grows out of a high-energy bound state,
indicating that some incoherent bands start to form coherent quasiparticles
with an underlying Fermi surface, which is the typical signature for the
incoherent-coherent crossover \cite{Takenaka2002,Stadler2015}.
Previous ARPES measurements~\cite{Miao2016,Yi2015} observed a slight decrease of
the spectral weight of a band near the Fermi level with increasing temperature.
Here, we offer clear evidence of spectral weight transfer during this process in
a system that is close to half filling.

Because D3's intensity is significantly enhanced and becomes very sharp 
at low temperatures, it dominates the DC conductivity below 75~K 
[Fig. S2(b) in the Supplementary Material]; its $T^2$-dependent scattering rate, 
indicating FL behavior~\cite{Dai2015}, is in agreement with recent transport 
measurements~\cite{Xiang2016}.  The ultra-low scattering rate for D3 reflects high
quality of these samples (RRR$\sim 512$); however, the presence of disorder might
explain the absence of FL behavior observed in another study~\cite{Dong2010}.
Comparing with D3, even though D1's contribution to the DC conductivity is very
small, the existence of two different kinds of narrow Drude components points to strong
orbital differentiation in FeSCs~\cite{Hardy2013,DeMedici2011}.

%
%
To better understand the origin of the emergent Drude component, calculations
for KFe$_2$As$_2$ were performed using density functional theory (DFT),
and extended using dynamical mean field theory (DMFT) (the details of which are
described in the Supplemental Material); DFT+DMFT has been shown to accurately
reproduce the electronic behavior of strongly correlated materials where
DFT alone typically fails \cite{Haule2010,Yin2011,Yin2011a}.
The results are summarized in Fig.~\ref{fig:dmft}.  Consistent with ARPES measurements,
only hole-like Fermi surfaces are present in KFe$_2$As$_2$ \cite{Yoshida2014,Sato2009}.
From the temperature-dependent band structure, we notice that at high temperature
[293~K, Fig.~\ref{fig:dmft}(a)], the band with $d_{xy}$ orbital character is much fainter
than those dominated by the $d_{xz/yz}$ orbitals.  However, at low temperature
[58~K, Fig.~\ref{fig:dmft}(b)] the $d_{xy}$ character increases dramatically.
These changes are reflected in the temperature dependence of the density-of-states
(DOS) for different orbital characters [Fig.~\ref{fig:dmft}(d) at 293~K, and
Fig.~\ref{fig:dmft}(e) at 58~K], where the $d_{xy}$ DOS sharpens at low temperature
and dominates the low-energy DOS, signalling an incoherent-coherent crossover.
This suggests that the emergent coherent peak in the optical conductivity below
150~K is most likely dominated by the $d_{xy}$ orbital.

In most iron-based superconductors, the state near $E_{\rm F}$ arises mainly from the $t_{2g}$ 
orbitals; correlation effects will give rise to charge transfer from the $d_{xy}$ to $d_{xz/yz}$
orbitals~\cite{Lee2012,Li2016}. Comparing the DFT+DMFT band structure in Fig.~\ref{fig:dmft}(b)
with that of DFT in \ref{fig:dmft}(c) [also see Fig.~S5(b) in the Supplemental Material],
the presence of electronic correlations leads to an increase in the contribution of the
$d_{xy}$ band to the Fermi surface, while the $d_{xz/yz}$ contribution decreases in order
to maintain the Luttinger count.  In this hole-overdoped system (5.5\emph{e}/Fe), charge
transfer pushes the $d_{xy}$ orbital much closer to half filling~\cite{DeMedici2014}, and
Hund's rule coupling results in a strong renormalization of this orbital~\cite{DeMedici2011,DeMedici2014},
with the renormalization factor $1/Z \propto U/t$, where $t$ is the hybridization magnitude
and $Z$ is the quasiparticle spectral weight. Thus, at high temperature the $d_{xy}$ orbital
is more incoherent (localized) and contributes to the local moment, resulting in a Curie-Weiss
susceptibility~\cite{Hardy2013}.  At low temperature, the quasiparticle spectral weight
(proportional to the hybridization of $d_{xy}$ orbital between nearest-neighbor atoms),
is enhanced continuously~\cite{Fobes2014} [Fig.~\ref{fig:dmft}(f)].  Below 150~K, this
orbital begins to delocalize and form coherent quasiparticles, which is consistent with
the spectral transfer from the high to low-energy region [Fig.~\ref{fig:fits}(c)] and
the formation of a new coherent peak in our optical conductivity. Below 75~K, FL 
behavior ($1/\tau_{D3}\propto T^2$) is observed.  This result, combined with the 
Pauli-like susceptibility~\cite{Hardy2013}, indicates Kondo-type
screening, during which the local moments are totally screened by the conduction
electrons, resulting in a saturated spin susceptibility and diminished scattering
from local spins~\cite{Stadler2015}.

The absence of this behavior in a simple DFT calculation indicates that this
incoherent-coherent crossover is the result of electronic correlations driven by
Hund's rule coupling~\cite{Fanfarillo2015,Haule2009} (Fig.~S5 in the Supplementary
Material).  While the spectral weight in the optical conductivity is transferred
from low to high-energy region upon warming, the absence of a gap-like structure in
the optical conductivity and the residual DOS of the $d_{xy}$ orbital near the Fermi
level at 293~K are inconsistent with the description of an orbital-selective Mott
transition~\cite{Qazilbash2007}; instead, our observation suggests an orbital-selective
incoherent-coherent crossover in a Hund's metal~\cite{Stadler2015,Yin2012,Mravlje2011}.

%
%

%
%
In summary, we have observed an emergent Drude peak in the optical conductivity of
KFe$_2$As$_2$ associated with the incoherent-coherent crossover below $T^\ast\simeq 155$~K,
and determined that it originates from a high-energy bound state. Below 75~K, this response
sharpens quickly and shows FL behavior, which may come from Kondo-type screening by the
delocalized electrons. Based on DFT+DMFT calculations, we find that this new peak is dominated
by the $d_{xy}$ orbital, reflecting an orbital-selective incoherent-coherent crossover.
We propose that in KFe$_2$As$_2$, which is close to half filling, the incoherent-coherent
crossover is related to Hund's rule driven Kondo-type screening, rather than orbital-selective
Mottness.

%
%

\begin{acknowledgments}
We thank I. Zaliznyak, A. Wang and W. Yin for useful discussions.
Work at Chinese Academy of Science was supported by NSFC (Project No. 11374345 and No. 91421304)
and  MOST (Project No. 2015CB921303 and No. 2015CB921102).
Work at Beijing Normal University was supported by the National Natural Science Foundation
of China (Grant No. 11674030) and the National Key Research and Development Program of China
(contract No. 2016YFA0302300).
Work at Brookhaven National Laboratory was supported by the Office of Science,
U.S. Department of Energy under Contract No. DE-SC0012704.
\end{acknowledgments}
%
%
%

%

\vfill
\eject

%
%

\pagenumbering{gobble}

\begin{figure}[t]
  \vspace*{-1.5cm}
  \hspace*{-1.80cm}
  \includegraphics[width=8.40in]{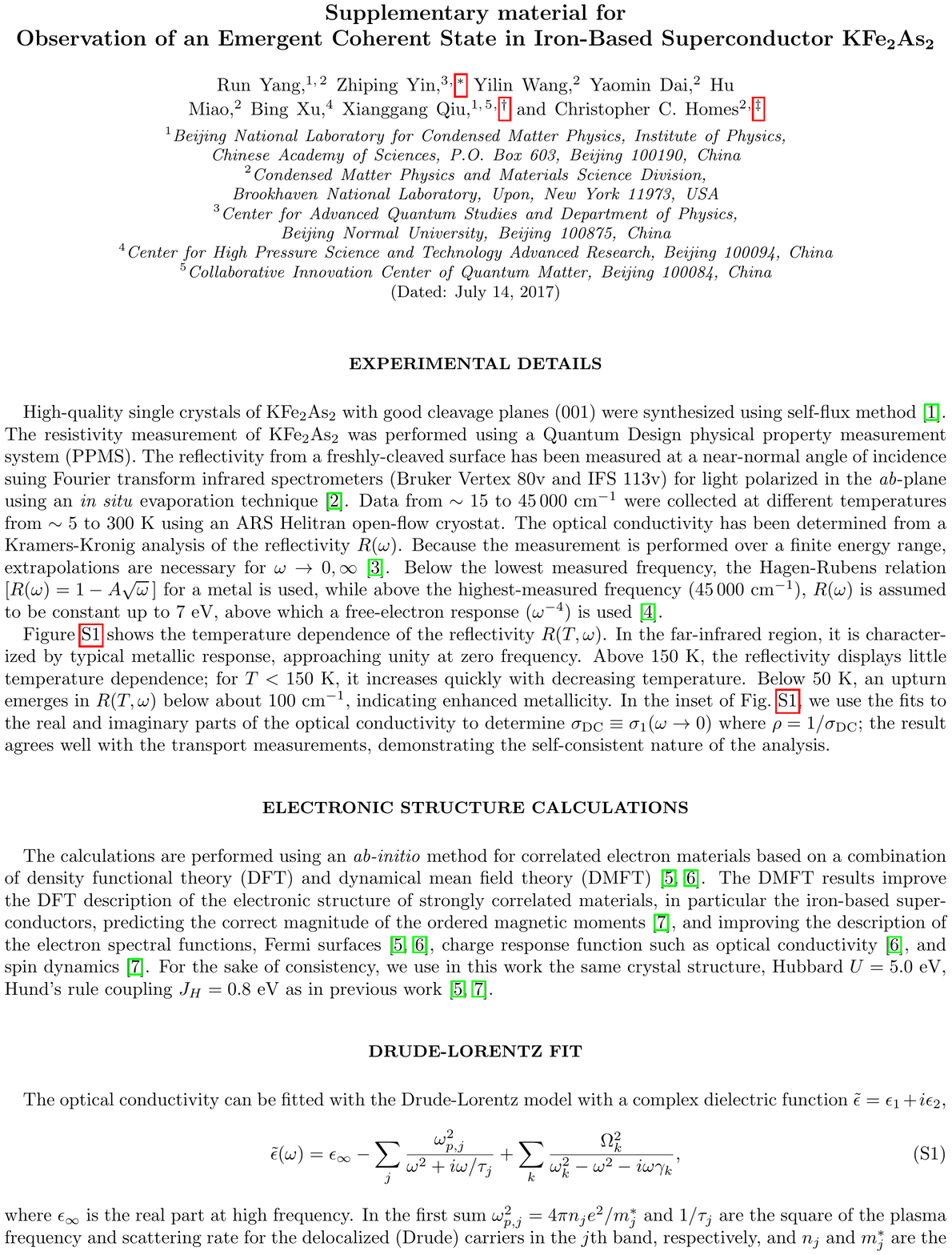}
\end{figure}

\begin{figure}[t]
  \vspace*{-1.5cm}
  \hspace*{-1.80cm}
  \includegraphics[width=8.40in]{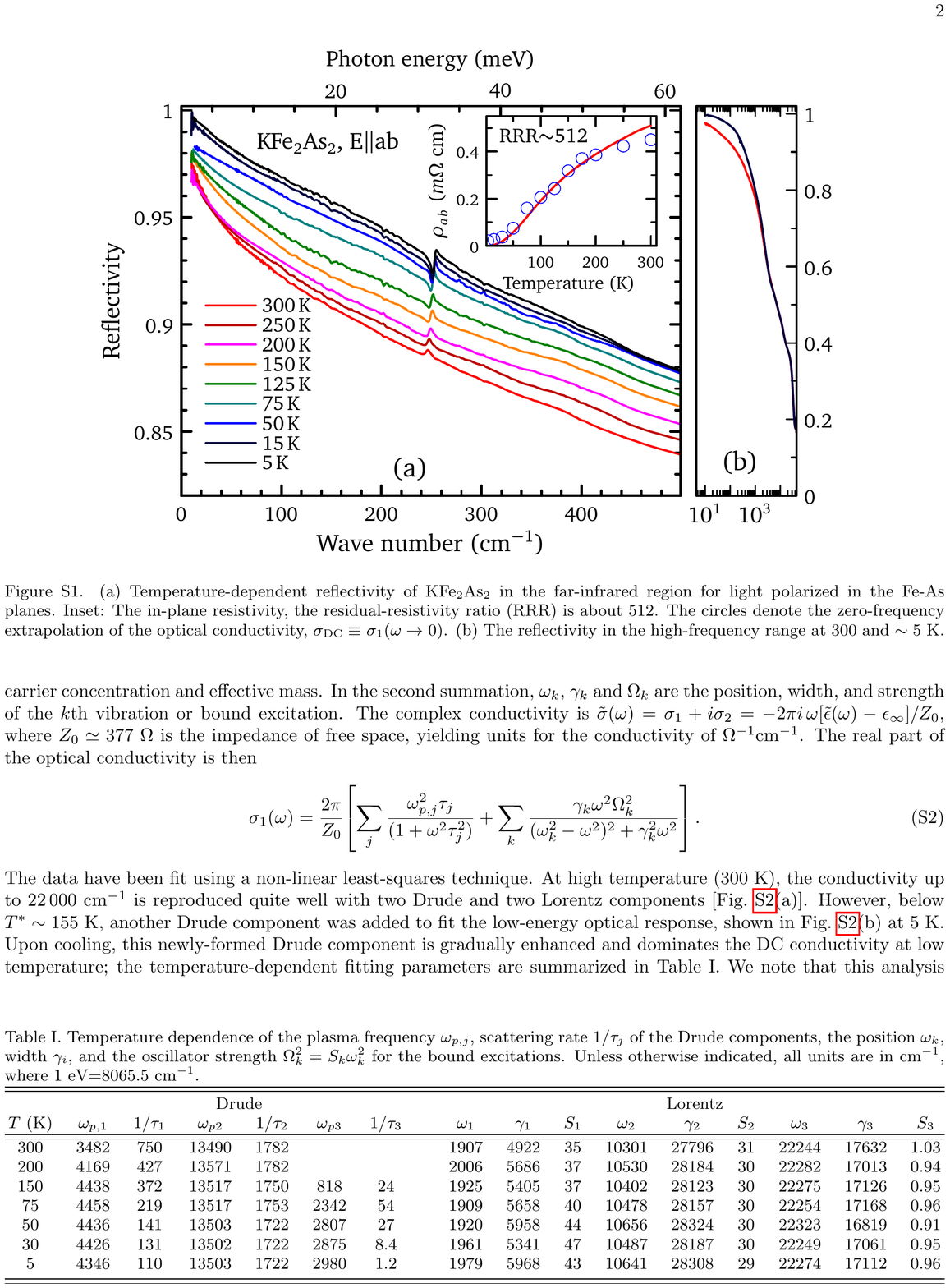}
\end{figure}

\begin{figure}[t]
  \vspace*{-1.5cm}
  \hspace*{-1.80cm}
  \includegraphics[width=8.40in]{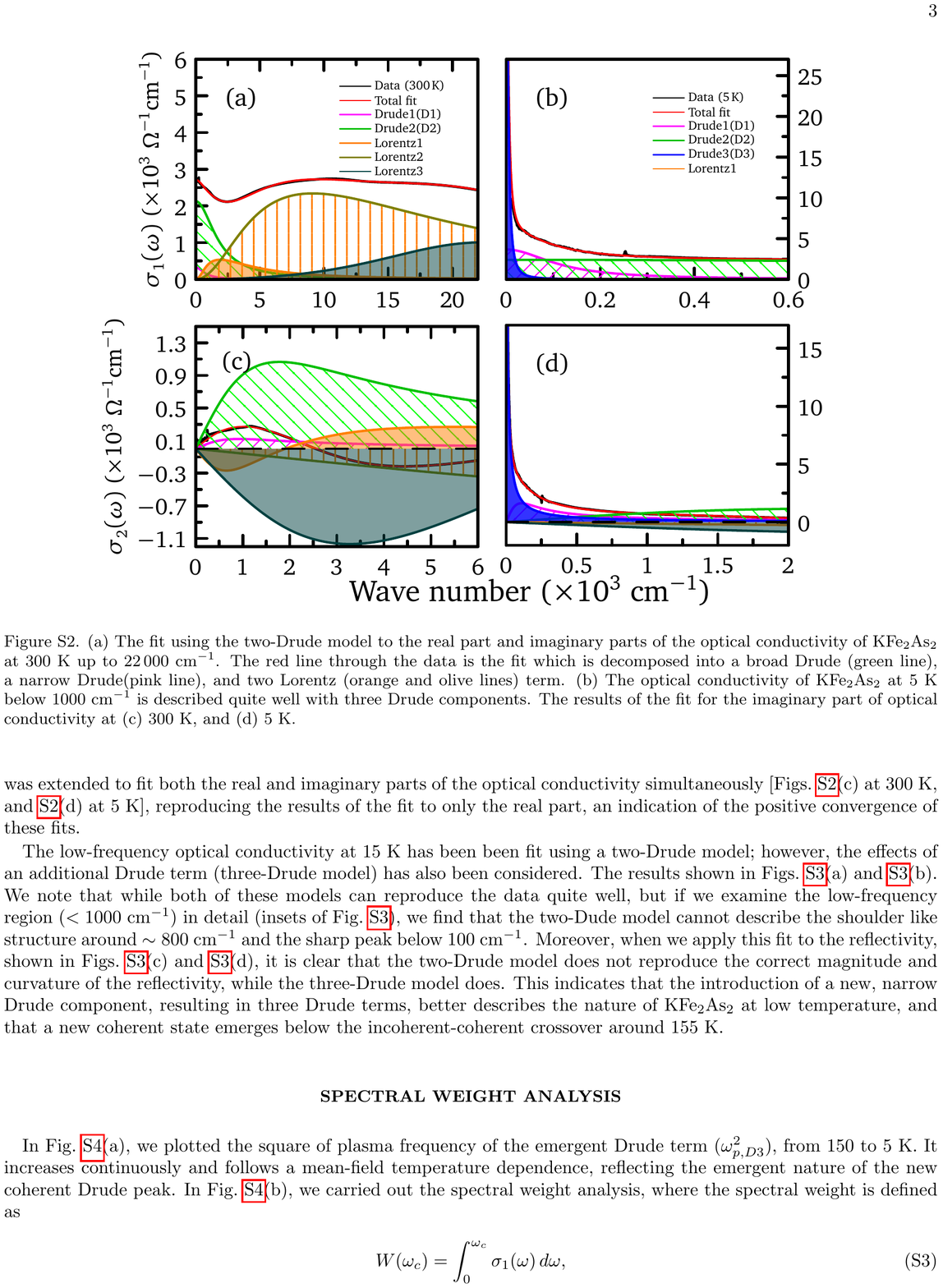}
\end{figure}

\begin{figure}[t]
  \vspace*{-1.5cm}
  \hspace*{-1.80cm}
  \includegraphics[width=8.40in]{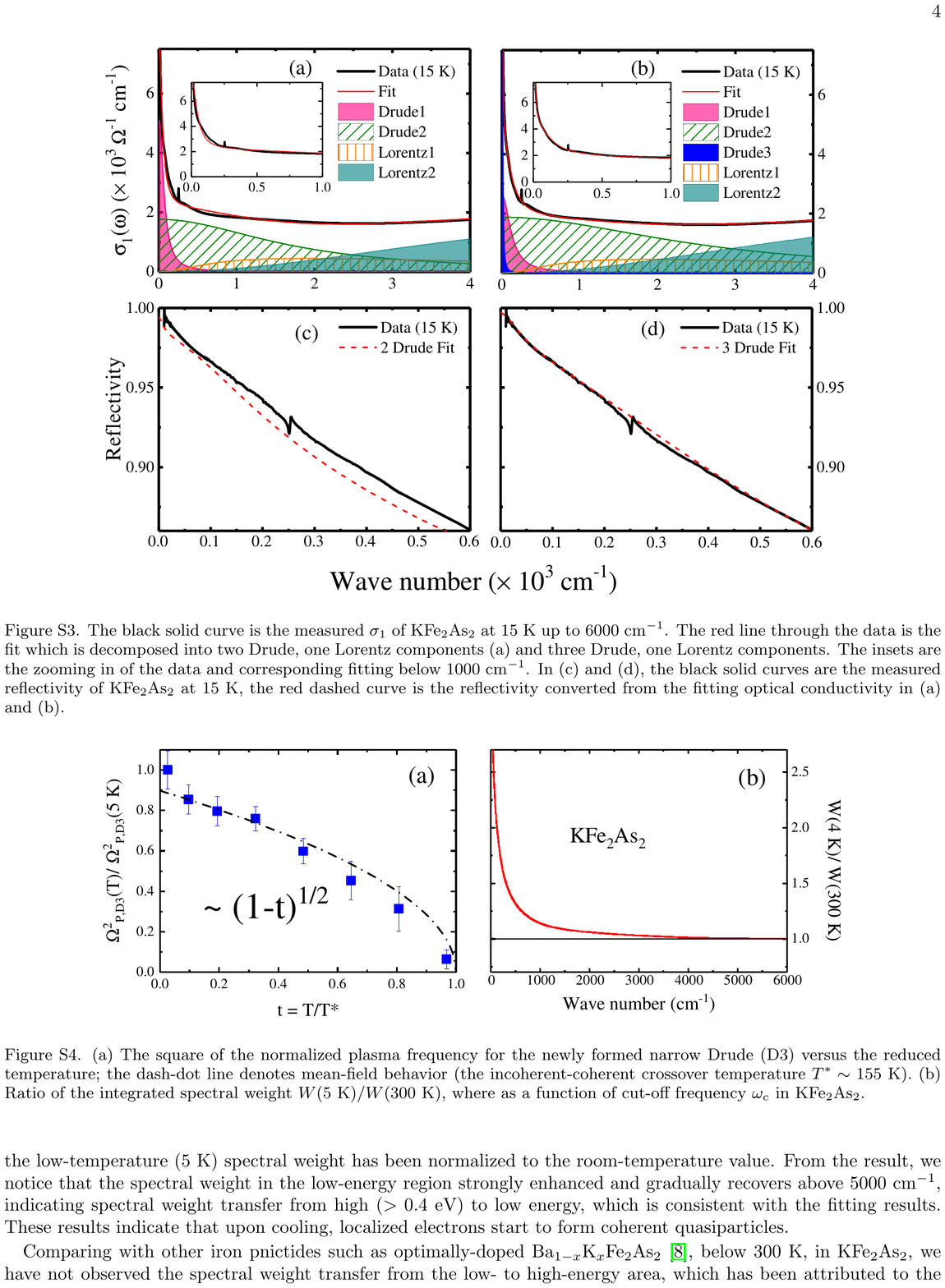}
\end{figure}

\begin{figure}[t]
  \vspace*{-1.5cm}
  \hspace*{-1.80cm}
  \includegraphics[width=8.40in]{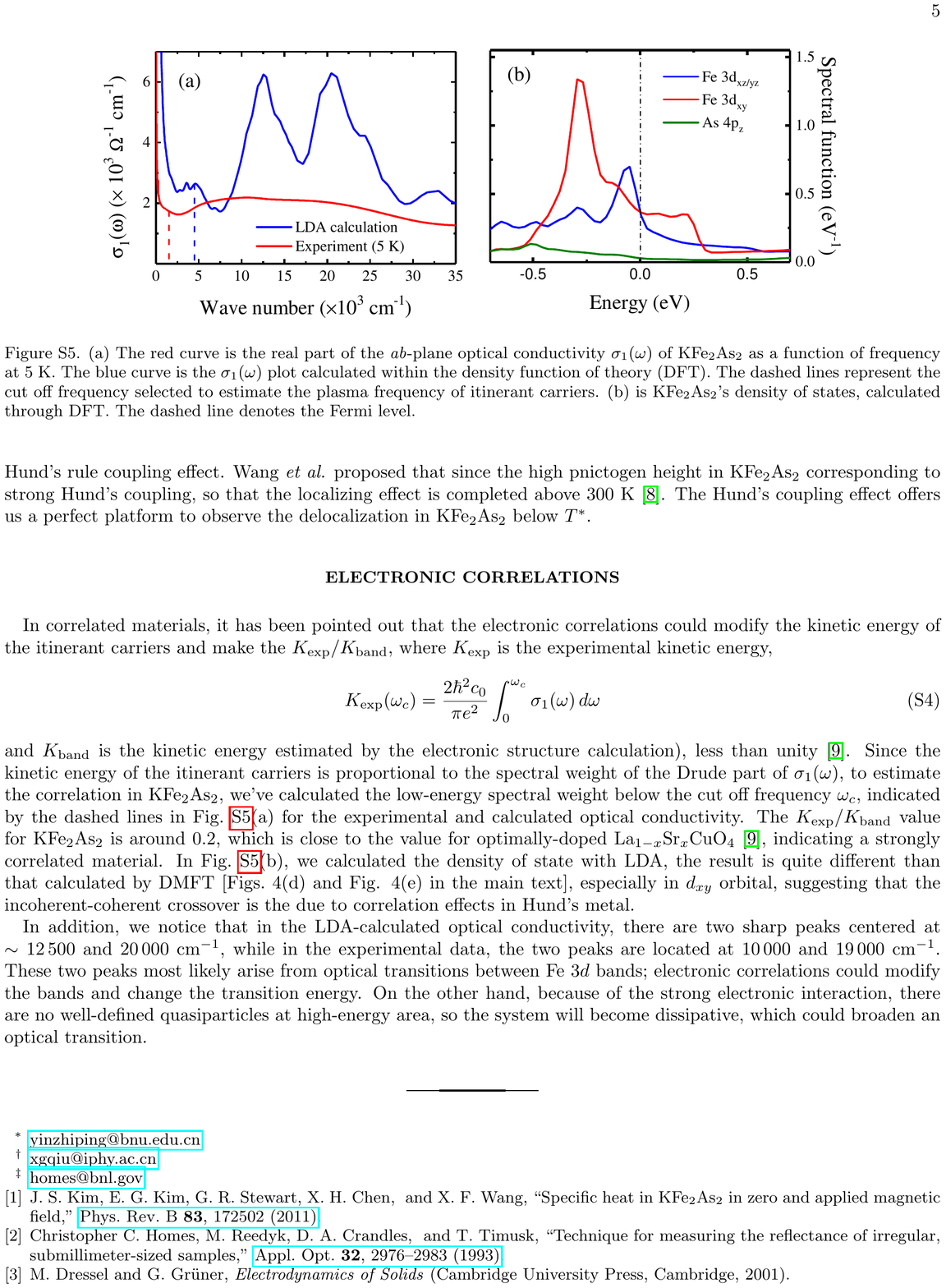}
\end{figure}

\begin{figure}[t]
  \vspace*{-1.5cm}
  \hspace*{-1.80cm}
  \includegraphics[width=8.40in]{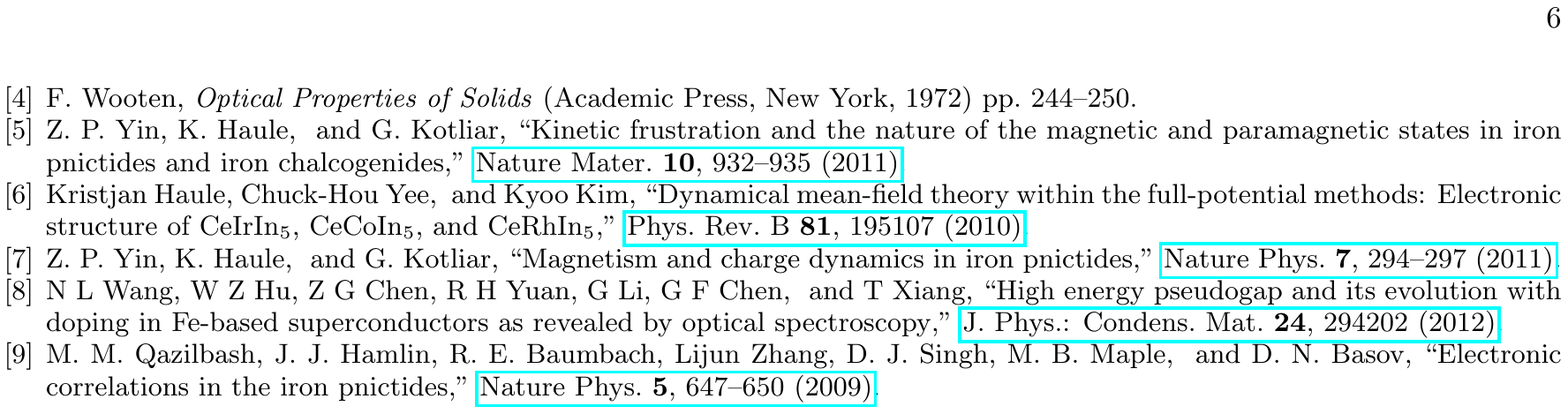}
\end{figure}

\end{document}